\documentclass{elsart}
\usepackage{amssymb,amsmath}
\def\equ#1{(\ref{#1})}
\def\cal{\mathcal}
\begin{document}
\begin{frontmatter}
\title{Generating mass and topological terms to the
antisymmetric tensor matter field by Higgs mechanism } 
\author[uece]{L. Gonzaga Filho, M. S. Cunha,} \author[ufc]{C. A. S. Almeida
and R. R. Landim\thanksref{renan}}
   \address[uece]{N\'ucleo de F\'{\i}sica, Universidade Estadual do Cear\'a -
Av. Paranjana, 1700, CEP 60740-000, Fortaleza, Cear\'a, Brazil}
\address[ufc]{Departamento de F\'{\i}sica, Universidade Federal do
Cear\'a - Caixa Postal 6030, CEP 60455-760, Fortaleza, Cear\'a, Brazil}
\thanks[renan]{renan@fisica.ufc.br}

\begin{abstract}
The interaction between the complex antisymmetric tensor matter
field and a scalar field is constructed. We analyze the Higgs
mechanism and show the generation of mass and topological terms by
spontaneous symmetry breaking.
\end{abstract}
\begin{keyword}
Spontaneous breaking of gauge symmetries 
\PACS 11.15.Ex \sep 11.30.Er
\end{keyword}
\end{frontmatter}
Antisymmetric tensor  matter (ATM) fields are new objects of study in field theory from theoretical
as phenomenological point of view~\cite{bol,aki,pob,chiz,chizbook,geyer}. They naturally arises in conformal
field theory and conformal supergravity~\cite{tod,wit}. When a ATM field is coupled to an abelian
gauge field, it gives an asymptotically free ultraviolet behavior to the gauge coupling,
\textit{i.e.}, the renormalization beta function becomes negative\cite{ac}, and the model is
renormalizable in all order of perturbation theory~\cite{rer}. Another remarkable fact is that the
model can be seen as a $\lambda\varphi^4$ theory of a complex antisymmetric tensor which obeys a
complex self-dual condition~\cite{phi4}.

Some aspects of theories which involve ATM fields still remain to be discussed and clarified. For
instance, in Euclidean space they describe three physical and three ghost degrees of
freedom~\cite{Frad}. The study of the classical dynamics in the Minkowski space has been done by
Avdeev and Chizhov~\cite{queer} where they have argued the Hamiltonian becomes positive-definite if
the solutions are restricted to that bounded at the time infinity. In this case, only two degrees
of freedom contribute to the energy and momentum.

As shown by Lemes {\it et al}~\cite{phi4}, another aspect of ATM fields forbids the existence of a
mass term to them, namely the complex self-dual condition. Despite of this, in this paper we
analyze the possibility to give mass to a ATM field through the Higgs mechanism. As will be shown,
a scalar field is coupled to the ATM field and by requirement of parity it is described as a
doublet where one of its components is a pseudo-scalar. On the other hand, neglecting the parity, a
topological term to the ATM field can also be generated by spontaneous symmetry breaking.

Let us begin  by introducing the notations and conventions that will be used throughout this paper.
We use the metric $\eta_{\mu\nu}={\rm diag}~(+,-,-,-)$, and the  totally antisymmetric tensor
$\varepsilon_{\mu\nu\rho\sigma}$ is normalized as $\varepsilon_{0123}=1$. In Minkowski space-time,
we take the antisymmetric tensor matter field $T_{\mu \nu }$ as real component of a complex second
rank tensor $\varphi _{\mu \nu }$ which obeys the complex self-dual condition~\cite{phi4}, namely
\begin{equation}
\varphi _{\mu \nu }=i\widetilde{\varphi }_{\mu \nu } \label{selfdual}
\end{equation}
where
\begin{equation}
\varphi _{\mu \nu }=T_{\mu \nu }+i\widetilde{T}_{\mu \nu }, \quad  \widetilde{\varphi }_{\mu \nu
}=\frac 12\varepsilon _{\mu \nu \rho \sigma }\varphi ^{\rho \sigma }. \label{eq2}
\end{equation}
Thus the complex ATM field can be coupled to axial Abelian gauge field and Dirac spinors in a more
compact way than that originally proposed by Avdeev and Chizhov~\cite{ac},

\begin{eqnarray}
 S_{inv} & = &  \int d^4x \left\{-\dfrac{1}{4g^2}F_{\mu\nu} F^{\mu\nu}+ i \bar\psi \gamma^{\mu} \partial_{\mu} \psi
 +h\bar\psi\gamma_{5}\gamma^{\mu}A_{\mu}\psi - (\nabla_{\mu} \varphi^{\mu\nu})^{\dagger}
      (\nabla^{\sigma} \varphi_{\sigma\nu}) \right. \nonumber\\
  &  &  \left. -\dfrac{q}{8} ( {\varphi^{\dagger\mu\nu}} \varphi_{\nu\alpha} {\varphi^{\dagger\alpha\beta}}
      \varphi_{\beta\mu}) + \frac{1}{2}y \bar\psi \sigma_{\mu\nu}(\varphi^{\dagger\mu\nu}+
      \varphi^{\mu\nu}) \psi \right\}\ . \label{L}
\end{eqnarray}

 \noindent  The above action is  invariant under the following gauge
 transformations,
\begin{eqnarray}
\nonumber&\delta A_\mu = \partial_\mu\omega \ ,&~~~~ \delta \psi = -ih\omega \gamma_{5} \psi \\
&\delta \bar\psi = -ih\omega \bar\psi \gamma_{5} \ ,&~~ \delta \varphi_{\mu\nu} = 2ih\omega
\varphi_{\mu\nu} \ .\label{gauge}
\end{eqnarray}

\noindent Under parity~($\mathcal{P}$) and charge conjugation~($\mathcal{C}$):

\vspace{4mm}

${\ }{\ }i)$ Parity $\mathcal{P}$
\begin{equation}\begin{array}{l}
 x \rightarrow  x_p=(x^0,-x^{i})  \ , \qquad i=1,2,3 \\ [2mm]
 \psi\rightarrow \psi^p = \gamma^0 \psi \ ,\\[2mm] A_0\rightarrow
 A_0^p = - A_0, ~~~~A_i\rightarrow A_i^p = A_i \ , \\[2mm]
 \varphi_{0i}\rightarrow  \varphi^p_{0i} = -  \varphi^{\dagger}_{0i},
  ~~\varphi_{ij} \rightarrow \varphi^p _{ij} = \varphi^{\dagger}_{ij}  \  .
\end{array}\label{parity}\end{equation}

\vspace{4mm} ${\ }{\ }ii)$ Charge conjugation $\mathcal{C}$
\begin{equation}\begin{array}{l}
  \psi \rightarrow \psi^{c} = C \bar{\psi}^{T} \ , \qquad
  C=i\gamma^{0} \gamma^{2} \ , \\[2mm] A_{\mu} \rightarrow A_{\mu}^{c}
  = A_{\mu}  \ , \\[2mm] \varphi_{\mu\nu} \rightarrow \varphi_{\mu\nu}^{c} = -
  \varphi_{\mu\nu}  \ .
\end{array}\label{c-conjg}\end{equation}


Now we construct the coupling between the complex self-dual field $\varphi_{\mu\nu}$ and the
complex scalar field $\phi=\phi_1+i\phi_2$ restricting ourselves to the power-counting
renormalizable interactions, gauge invariance, and the parity symmetry of the model described in
\equ{gauge} and \equ{parity}, respectively.

Since we are interested in mass generation, we take only quadratic terms in $\varphi_{\mu\nu}$.
However, from the self-dual condition \equ{selfdual} and the properties of the Levi-Civita tensor
in Minkowsky space-time we have $\varphi_{\mu\nu}^*\varphi^{\mu\nu}=0$, which implies that the most
general quadratic terms in $\varphi_{\mu\nu}$  coupled to a complex scalar $\phi$ and
renormalizable by power-counting  are the form
\begin{equation}\label{interac}
  \int \left(d^4x~ a\, \varphi^{*\mu\nu}\varphi_{\mu\nu}^*\phi+b\,
  \varphi^{\mu\nu}\varphi_{\mu\nu}\phi+c\,
  \varphi^{*\mu\nu}\varphi_{\mu\nu}^*\phi\phi+d\,
  \varphi^{\mu\nu}\varphi_{\mu\nu}\phi\phi+{\rm c.c}\right),
\end{equation}
where  c.c stands for complex conjugate terms such that \equ{interac} becomes real with $a,b,c$,
and $d$ arbitrary constants. Notice that, from the action given by Eq.~\equ{L}, the self-dual
tensor $\varphi_{\mu\nu}$ has canonical dimension equal to one. The value is the same for a scalar
field in four dimensions.

The requirement of gauge invariance has given us four different gauge transformations for the
scalar field, with charges $\pm 2h$ and $\pm 4h$. This implies that only one term of ~\equ{interac}
can be added to the Lagrangian~\equ{L}.

Now turning back to the parity invariance of the interactions, the  quadratic term
 $\varphi_{\mu\nu}\varphi^{\mu\nu}$ is transformed under parity by
\begin{equation}\label{eq8}
  \varphi_{\mu\nu}\varphi^{\mu\nu}\rightarrow\varphi^*_{\mu\nu}\varphi^{*\mu\nu}.
\end{equation}
To make all the terms in \equ{interac} invariant under parity, $a, b, c$, and $ d$ must be real and
the complex scalar field $\phi$ must be transformed as $\phi\rightarrow\phi^*$. This implies that
$\phi_1\rightarrow\phi_1$ and $\phi_2\rightarrow-\phi_2$, \textit{i.e.},  $\phi_2$ is a
pseudo-scalar.

We are able now to analyze the spontaneous symmetry breaking, first focusing our attention to the
interaction term given in \equ{interac} with the parity being preserved. The Higgs potential is
\begin{equation}\label{eq9}
  V(\phi)=-\frac{1}{2}\mu^2\phi^*\phi+\frac{\lambda}{4}(\phi^*\phi)^2.
\end{equation}
The minimum of \equ{eq9} occurs at $\phi_1^2+\phi_2^2=v^2$, where $v=(\mu^2/\lambda)^{1/2}.$ In principle there exist  infinity number of physical equivalent minima.  Let us emphasize here that in   our case the  complex scalar field has a component which is not a scalar but a pseudo-scalar. Under parity the $\phi_2$ component is transformed as $\phi_2\rightarrow-\phi_2$. Consequently, the invariance of the model by parity symmetry fixes  its vacuum expected value to be zero, i.e., $\langle0|\phi_2|0\rangle=0$. Instead of infinity number of physical equivalent vacua we have only two possibilities: 
  $\langle0|\phi_2|0\rangle=0$ and $\langle0|\phi_1|0\rangle=v$ or $\langle0|\phi_2|0\rangle=0$ and $\langle0|\phi_1|0\rangle=-v$. According to the charge of $\phi$ field and redefining  the $\phi_1$ 
  field as $\phi_1=\phi_1'\pm{v}$, the interaction terms in \equ{interac} can be written in terms of $T_{\mu\nu}$ as:

\vspace{4mm} ${\ }{\ }i)$ $\phi$ with charge $+4h$
\begin{equation}\label{eq10}
  \int d^4x\,4a\,(T^{\mu\nu}\widetilde{T}_{\mu\nu}\phi_2+T^{\mu\nu}T_{\mu\nu}\phi_1'\pm{v}T^{\mu\nu}T_{\mu\nu}).
\end{equation}

\vspace{4mm} ${\ }{\ }ii)$ $\phi$ with charge $-4h$
\begin{equation}\label{eq11}
  \int d^4x\,4b\,(T^{\mu\nu}T_{\mu\nu}\phi_1'\pm{v}T^{\mu\nu}T_{\mu\nu}-T^{\mu\nu}\widetilde{T}_{\mu\nu}\phi_2).
\end{equation}

\vspace{4mm} ${\ }{\ }iii)$ $\phi$ with charge $+2h$
$$
   \int d^4x\,4c\,(({\phi_1'}^2-{\phi_2}^2)T^{\mu\nu}T_{\mu\nu}+2\pm{v}T^{\mu\nu}T_{\mu\nu}\phi_1'+v^2T^{\mu\nu}T_{\mu\nu})
$$
\begin{equation}\label{eq12}
+\int
d^4x\,4c\,(2\pm{v}T^{\mu\nu}\widetilde{T}_{\mu\nu}\phi_2+2T^{\mu\nu}\widetilde{T}_{\mu\nu}\phi_1'\phi_2)
\end{equation}

\vspace{4mm} ${\ }{\ }iv)$ $\phi$ with charge $-2h$
$$
   \int d^4x\,4d\,(({\phi_1'}^2-{\phi_2}^2)T^{\mu\nu}T_{\mu\nu}+2\pm{v}T^{\mu\nu}T_{\mu\nu}\phi_1'+v^2T^{\mu\nu}T_{\mu\nu})
$$
\begin{equation}\label{eq13}
-\int
d^4x\,4d\,(2\pm{v}T^{\mu\nu}\widetilde{T}_{\mu\nu}\phi_2+2T^{\mu\nu}\widetilde{T}_{\mu\nu}\phi_1'\phi_2)
\end{equation}
Let us examine carefully each type of interactions above. For a fixed value of the parameter $a$ there is no way to avoid tachyons because $\pm{v}a$ is not always positive. The same occur for interaction $ii)$. Then the first two types of interactions are not physically  acceptable. 
The interactions terms $(iii)$ and $(iv)$ have the mass term dependent on $v^2$ which is always positive. In order to avoid tachyons $c$ or  $d$ must be positive depending of the charge of $\phi$. Consequently the ATM field  acquire mass $m=4|v|\sqrt{c}$ for $\phi$ with charge $2h$ and mass $m=4|v|\sqrt{d}$ for $\phi$ with charge $-2h$. Each pole of ATM field is free of tachyon since we have only two allowed vacua compatible with the parity symmetry.

Let us now examine the case when the parity invariance of the theory is relaxed. In this case
 $\phi_1$ and $\phi_2$ are suppose to be scalars. To simplify the analysis, we
 consider $a,b,c$ and $d$ real constants  in \equ{interac}. We have
 now that the vacuum expected value of $\phi_2$ can be non  zero value, \textit{i.e.}, $\langle0|\phi_2|0\rangle=v_2$ and
 $\langle0|\phi_1|0\rangle=v_1$, where $v_1^2+v_2^2=\mu^2/\lambda$. This
 modify all the terms $(i)-(iv)$. For instance, $(i)$ reads
 \begin{equation}\label{eq14}
  \int d^4x\,4a\,(T^{\mu\nu}\widetilde{T}_{\mu\nu}\phi_2'+T^{\mu\nu}T_{\mu\nu}\phi_1'+v_1T^{\mu\nu}T_{\mu\nu}+v_2T^{\mu\nu}\widetilde{T}_{\mu\nu}).
\end{equation}
Besides a mass term, we also have a topological term given by
\begin{equation}\label{eq15}
 \int d^4x\,4av_2T^{\mu\nu}\widetilde{T}_{\mu\nu}.
\end{equation}
This is topological in a sense of metric independence, \textit{i.e.},
\begin{equation}\label{eq16}
  \int_{\cal{M}}T_{\mu\nu}\widetilde{T}^{\mu\nu}\,\sqrt{|g|}\,d^4x=-
  \frac{1}{2}\int_{\cal{M}}\varepsilon^{\mu\nu\alpha\beta}T_{\mu\nu}T_{\alpha\beta}\,\,d^4x,
\end{equation}
where ${\cal{M}}$ is a Lorentzian manifold. A similar mechanism occur in three dimensions with
nonminimal coupling, where the Chern-Simons term is generated by spontaneous symmetry breaking
through the covariant derivative \cite{lat}. Note that with the parity being relaxed the theory could presents tachyonic states.

Our study of mass generation for ATM fields is motivated by the interest in possible tensor interactions in weak decays. This is
connected with the  experiments on $\pi^-\longrightarrow e^-\tilde{\nu}\gamma$  and $ K^+\longrightarrow\pi^0e^+\nu$ decays \cite{aki,bol}. The experimentally obtained form factors cannot be explained in the
framework of the standard electroweak theory.  In Ref. \cite{pob}
results of the experiment on $\pi^-\longrightarrow e^-\tilde{\nu}\gamma$ decay were explained by introducing an additional tensor interaction in the Fermi Lagrangian.

Summarizing, in this letter we have shown that the Higgs mechanism
gave mass to the ATM field even though complex ATM field
$\varphi_{\mu\nu}$ cannot have such a explicit mass term. Through the Higgs mechanism, the parity symmetry allows only two possible physically acceptable types of interaction. Another
interesting result came from relaxing the obligation of parity
invariance of the complex scalar field. In this case, we could
obtain massive and topological terms by spontaneous symmetry
breaking (SSB).
Differently of Ref. [12] where the topological Chern-Simons term
generation occurred through gauge covariant derivative, here by SSB
it was possible to produce a topological term when we coupled the
complex ATM field with a scalar field. Our work can be compared with
\cite{gonzaga} where  we propose a mechanism to generate mass to
ATM field which preserves the U(1) symmetry. By this
mechanism, a topological term is introduced via a complex vector field.

This work was supported by CNPq (Brazilian Research Agency) and Funda\c{c}\~ao Cearense de Apoio ao
Desenvolvimento Cient\'{\i}\-fico e Tecnol\'ogico do Estado do Cear\'a - FUNCAP.

\end{document}